\documentstyle[11pt,psfig,aaspp4]{article}

\begin{document}

\title{Extragalactic Foregrounds of the Cosmic Microwave
Background: Prospects for the MAP Mission}

\author{Alexandre Refregier\altaffilmark{1}, David N. Spergel}
\affil{Department of Astrophysical Sciences, Princeton University,
Princeton, NJ 08544}
\and
\author{Thomas Herbig}
\affil{Physics Department, Princeton University, Jadwin Hall,
P.O. Box 708, Princeton, NJ 08544}
\altaffiltext{1}{email: refreg@astro.princeton.edu}

\begin{abstract}
While the major contribution to the Cosmic Microwave Background (CMB)
anisotropies are the sought-after primordial fluctuations produced at
the surface of last scattering, other effects produce secondary
fluctuations at lower redshifts.  Here, we study the extragalactic
foregrounds of the CMB in the context of the upcoming \hbox{MAP}
mission. We first survey the major extragalactic foregrounds and show
that discrete sources, the Sunyaev-Zel'dovich (SZ) effect, and
gravitational lensing are the most dominant ones for \hbox{MAP}. We
then show that \hbox{MAP} is expected to detect ($>5\sigma$) about 46
discrete sources directly with 94 GHz fluxes above 2 Jy. The most
prominent SZ features on the CMB sky are rich clusters of galaxies. In
particular, we show that the Coma cluster will be clearly detected and
marginally resolved by \hbox{MAP}. We then consider a cosmological
population of clusters, and show that \hbox{MAP} should detect
($>5\sigma$) about 10 SZ clusters directly. The mean SZ fluxes of
fainter clusters can be probed by cross-correlating \hbox{MAP} with
cluster positions extracted from existing catalogs. For instance, a
\hbox{MAP}-XBACs cross-correlation will be sensitive to clusters with
$S(94{\rm GHz}) \gtrsim 200$ mJy, and will thus provide a test of
their virialization state and a measurement of their gas fraction.
Finally, we consider probing the hot gas on supercluster scales by
cross-correlating the CMB with galaxy catalogs. Assuming that galaxies
trace the gas, we show that a cross-correlation between \hbox{MAP} and
the APM catalog should yield a marginal detection, or at least a
four-fold improvement on the COBE upper limits for the {\it rms}
Compton $y$-parameter.
\end{abstract}


\section{Introduction}
The Cosmic Microwave Background (CMB) provides a unique probe to the
early universe (see \markcite{whi94}White, Scott, \& Silk 1994 for a
review). If CMB fluctuations are consistent with inflationary models,
future ground-based and satellite experiments will yield accurate
measurements of most cosmological parameters (see
\markcite{zal97}Zaldarriaga, Spergel, \& Seljak 1997;
\markcite{bond97}Bond, Efstathiou, \& Tegmark 1997 and reference
therein). These measurements rely on the detection of primordial
anisotropies produced at the surface of last scattering. However,
various secondary effects produce fluctuations at lower redshifts. The
study of these secondary fluctuations (or extragalactic foregrounds)
is important in order to isolate primordial fluctuations.  In
addition, secondary fluctuations are interesting in their own right
since they provide a wealth of information on the local universe.

The upcoming \hbox{MAP} mission (\markcite{ben95}Bennett et al.\ 1995)
will produce a map of the CMB sky with unprecedented sensitivity,
frequency coverage, and angular resolution. In this paper, we study
the extragalactic foregrounds of the CMB in the context of \hbox{MAP}.
After surveying the major sources of secondary anisotropies, we show
that discrete sources, the Sunyaev-Zel'dovich (SZ) effect, and
gravitational lensing are the dominant extragalactic foregrounds for
\hbox{MAP}. The SZ effect is produced by the comptonization of CMB
photons by the hot gas in clusters and superclusters of galaxies along
the line of sight. Searches for extragalactic foregrounds have been
performed in the COBE maps, but have only led to upper limits
(\markcite{bou93}Boughn \& Jahoda 1993; \markcite{ben93}Bennett et
al.\ 1993; \markcite{ban96}Banday et al.\ 1996;
\markcite{kne97}Kneissl et al.\ 1997). We focus on discrete and SZ
foregrounds and show how they affect future CMB maps.  In addition, we
present statistical techniques to detect SZ fluctuations beyond the
threshold of direct detection for \hbox{MAP}, by cross-correlating CMB
maps with existing galaxy and cluster catalogs. While the present
analysis was developed in the context of \hbox{MAP}, many of the
techniques will also be useful for other experiments (esp. Planck
Surveyor; \markcite{ber96}Bersanelli et al. 1996).

In \S\ref{MAP}, we briefly describe the characteristics of \hbox{MAP} which
are relevant to the present study. In \S\ref{foregrounds}, we review
the major sources of secondary anisotropies, and quantify their
relative importance for \hbox{MAP}. In \S\ref{discrete}, we estimate, using
existing models, the number of discrete sources detectable with
\hbox{MAP}. In \S\ref{SZ}, we turn to the contribution of the SZ Effect to
CMB maps. We first survey the most prominent SZ sources in the sky
(\S\ref{SZ_prominent}), and then consider the SZ effect by a
cosmological population of clusters (\S\ref{SZ_clusters}), and by
superclusters (\S\ref{SZ_superclusters}). Finally, \S\ref{conclusion}
summarizes our conclusions.

\section{\hbox{MAP} Sensitivity}
\label{MAP}
The \hbox{MAP} instrument will comprise 5 channels with frequencies
between 20 and 94 GHz. A detailed description of the \hbox{MAP}
mission can be found in \markcite{ben95}Bennett et al.\
1995. Foregrounds can be optimally separated using multi-frequency
filtering techniques (\markcite{ber96}Bersanelli et al. 1996;
\markcite{teg96}Tegmark \& Efstathiou 1996; \markcite{bou98}Bouchet et
al. 1998). As we will see below, extragalactic foregrounds are, for
the most part, only weakly dependent on frequency in the MAP frequency
range. (Even for discrete sources, the improvement yielded by a
multi-channel analysis is limited by the presence of flat-spectrum
sources; see \S\ref{flat_spectrum}). As a first step and to be
conservative, we thus presently only consider the W-band channel which
has the highest angular resolution.

Assuming the current design with 2 years of coverage, the combined
W-band detectors have a center frequency of $\nu \simeq 94$ GHz, a
beam FWHM of $\theta_{\rm beam} \simeq 0.21$ deg, and an instrumental
{\it rms} noise of $\Delta T_{\rm pix} \simeq 35 \mu$K (thermodynamic
temperature) per $\Omega_{\rm pix}= \theta_{\rm pix}^{2} = 0\fdg3
\times 0\fdg3$ pixels.

\begin{figure}
\plotone{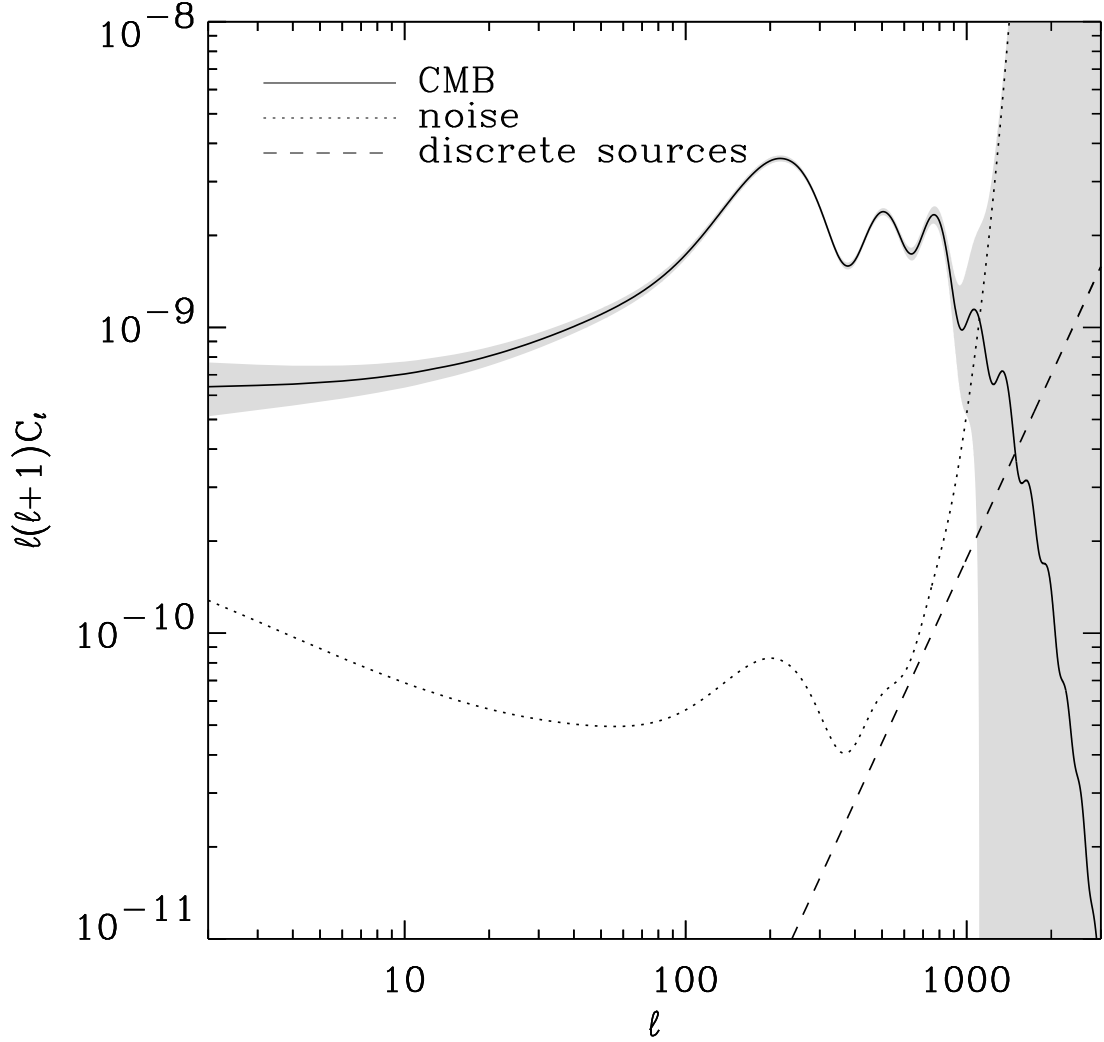}
\caption{Power spectrum sensitivity for the 94 GHz \hbox{MAP} channel.  The
solid line corresponds to a COBE normalized standard CDM model with
$\Omega_{b}=0.05$ and $h=0.5$. The shaded region and the dotted line
show the {\it rms} sensitivity for the \hbox{MAP} 94 GHz channel with a 2 year,
full sky coverage and a bandpass average of $\Delta \ell = 10$.  The
dashed line shows the power spectrum for residual discrete sources
with $S(94 {\rm GHz}) < S_{94}(5 \sigma) \simeq 2$ Jy, as extrapolated
from Toffolatti et al.\ (1998).
\label{fig:cl_disc}}
\end{figure}

The temperature fluctuations measured by \hbox{MAP} are decomposed
into the usual spherical harmonics basis, $\frac{\delta
T}{T_{0}}(\mbox{\bf $\theta$}) = \sum_{\ell,m} a_{\ell m} Y_{\ell
m}(\mbox{\bf $\theta$})$, where the $a_{\ell m}$'s are the multipole
moments. Here, $T_{0} \simeq 2.725$K is the mean CMB temperature
(\markcite{mat98}Mather et al.\ 1998). As a reference, the averaged
multipole moments $C_{\ell} \equiv \langle |a_{\ell m}|^{2} \rangle$
for a COBE-normalized standard CDM model (with $\Omega_b=0.05$, and
$h=0.5$) calculated with CMBFAST (\markcite{seljz96}Seljak \&
Zaldarriaga 1996) are shown in figure~\ref{fig:cl_disc}.
\label{wchan_param}

The {\it rms} uncertainty $\Delta C_{\ell}$ for measuring $C_{\ell}$
averaged over a bandpass of width $\Delta \ell$ is given by
(eg. \markcite{kno95}Knox 1995)
\begin{equation}
\Delta C_{\ell} = \left(\frac{2}{2 \ell +1} \right)^{\frac{1}{2}}
\left[ C_{\ell}+\left( \frac{\Delta T_{\rm pix}}{T_{0}} \right)^{2}
\Omega_{\rm pix} W_{\ell}^{-1} \right] (f_{\rm sky} \Delta \ell)^{-\frac{1}{2}}
\label{eq:delta_cl}
\end{equation}
where $\Delta T_{\rm pix}$ is the thermodynamic noise per pixel,
$\Omega_{\rm pix}$ is the pixel solid angle, and $f_{\rm sky}$ is the
fraction of the sky covered (with randomly-positioned discarded
pixels). Taking the beam to be gaussian with a standard deviation of
$\sigma_{\rm beam}$, the window function $W_{\ell}$ is given by
$W_{\ell}=e^{-\ell^2 \sigma_{\rm beam}^2}$. The first and second terms
correspond to cosmic variance and instrumental noise,
respectively. The measurement uncertainty $\Delta C_{\ell}$ for the
\hbox{MAP} 94 GHz channel is shown in figure~\ref{fig:cl_disc} both as
the shaded regions and as the dotted line. A 2-year coverage of the
full sky and a bandpass width of $\Delta \ell = 10$ was assumed.

\placefigure{fig:cl_disc}

The zero-lag temperature variance is given by 
\begin{equation}
\label{eq:deltaT_cmb}
(\Delta T)^{2} = \sum_{\ell} \frac{2 \ell +1}{4 \pi} C_{\ell}
W_{\ell} F_{\ell}T_{0}^{2} ,
\end{equation}
where $W_{\ell}$ is the beam window function defined above.  The
factor $F_{\ell}$ is an optional filter function (see below) and
should be set to 1 in the unfiltered case.  For the above CDM model ,
the unfiltered {\it rms} CMB fluctuation smoothed with the W-band beam
is $\Delta T_{\rm CMB} \simeq 92 \mu$K. The smoothed {\it rms}
fluctuation $\Delta T_{\rm inst}$, produced by the instrumental noise,
is related to the pixel {\it rms} noise $\Delta T_{\rm pix}$ by
$\Delta T_{\rm inst} = (4 \pi)^{-\frac{1}{2}} \theta_{\rm pix}
\sigma_{\rm beam}^{-1} \Delta T_{\rm pix}$. For the \hbox{MAP} 94 GHz
channel, the smoothed (unfiltered) instrumental noise is thus
$\Delta T_{\rm inst} \simeq 33 \mu$K.

For the detection of spatially compact sources, the CMB fluctuations
can be reduced by filtering out low spatial frequencies. Specifically,
we choose a gaussian filter function of the form $F_{\ell} = 1 -
e^{-\ell (\ell +1) \sigma^{2}_{\rm filter}}$ (see
\markcite{tego97}Tegmark \& Oliveira-Costa 1997, for a more
sophisticated treatment). For a filter with FWHM $\theta_{\rm
filter}\equiv \sigma_{\rm filter} \sqrt{8 \ln 2} =1^{\circ}$, the
resulting {\it rms} fluctuation is $\Delta T_{\rm CMB,filtered} \simeq
65 \mu$K, for the same model. Upon filtering, the instrumental noise
{\it rms} fluctuation becomes $\Delta T_{\rm inst,filtered} \simeq 32
\mu$K and is thus hardly affected. The resulting total {\it rms}
``noise'' produced by the combination of CMB and instrumental noise is
then
\begin{equation}
\label{eq:deltaT_noise}
\Delta T_{\rm noise} \simeq 73 ~\mu {\rm K}.
\end{equation}
We will use this figure when considering the detection of foreground
sources.

For future reference, we compute the sensitivity of \hbox{MAP} for
point source detections. For a 94 GHz map smoothed with a gaussian
beam, the peak temperature offset produced by an unresolved source with
flux $S(94 {\rm GHz}) \equiv S_{94}$ is
\begin{equation}
\Delta T_{94} \simeq 301 \left( \frac{0.21 {\rm deg}}{\theta_{\rm beam}}
\right)^{2} \left( \frac{S_{94}}{1 {\rm Jy}} \right) \mu{\rm K}.
\end{equation}
The corresponding $1\sigma$ threshold for the detection of point
sources is thus
\begin{equation}
\label{eq:s_onesigma}
S_{94}(1\sigma) \simeq 250 \left( \frac{\Delta T_{\rm noise}}{75 \mu
{\rm K}} \right) {\rm mJy},
\end{equation}
where the central value for $\Delta T_{\rm noise}$ was chosen to be
close to that of equation~(\ref{eq:deltaT_noise}).

\section{Overview of Extragalactic Foregrounds}
\label{foregrounds}
In this section, we survey the major extragalactic foregrounds and
secondary anisotropies of the CMB by relying on existing literature.
To assess their relative importance, we follow \markcite{teg96}Tegmark
\& Efstathiou (1996) and consider the quantity $ \Delta T_{\ell}
\equiv \left[ \ell(2 \ell+1) C_{\ell}/4 \pi \right]^{\frac{1}{2}} T_{0}$,
which gives the {\it rms} temperature fluctuations per $\ln \ell$ interval
centered at $\ell$. Another useful quantity that they considered is
the value of $\ell=\ell_{eq}$ for which foreground fluctuations are
equal to the CMB fluctuations, i.e.\ for which $C_{\ell}^{\rm
foreground} \simeq C_{\ell}^{\rm CMB}$. Note that, since the
foregrounds do not necessarily have a thermal spectrum, $\Delta
T_{\ell}$ and $\ell_{eq}$ generally depend on frequency.

\begin{figure}
\plotone{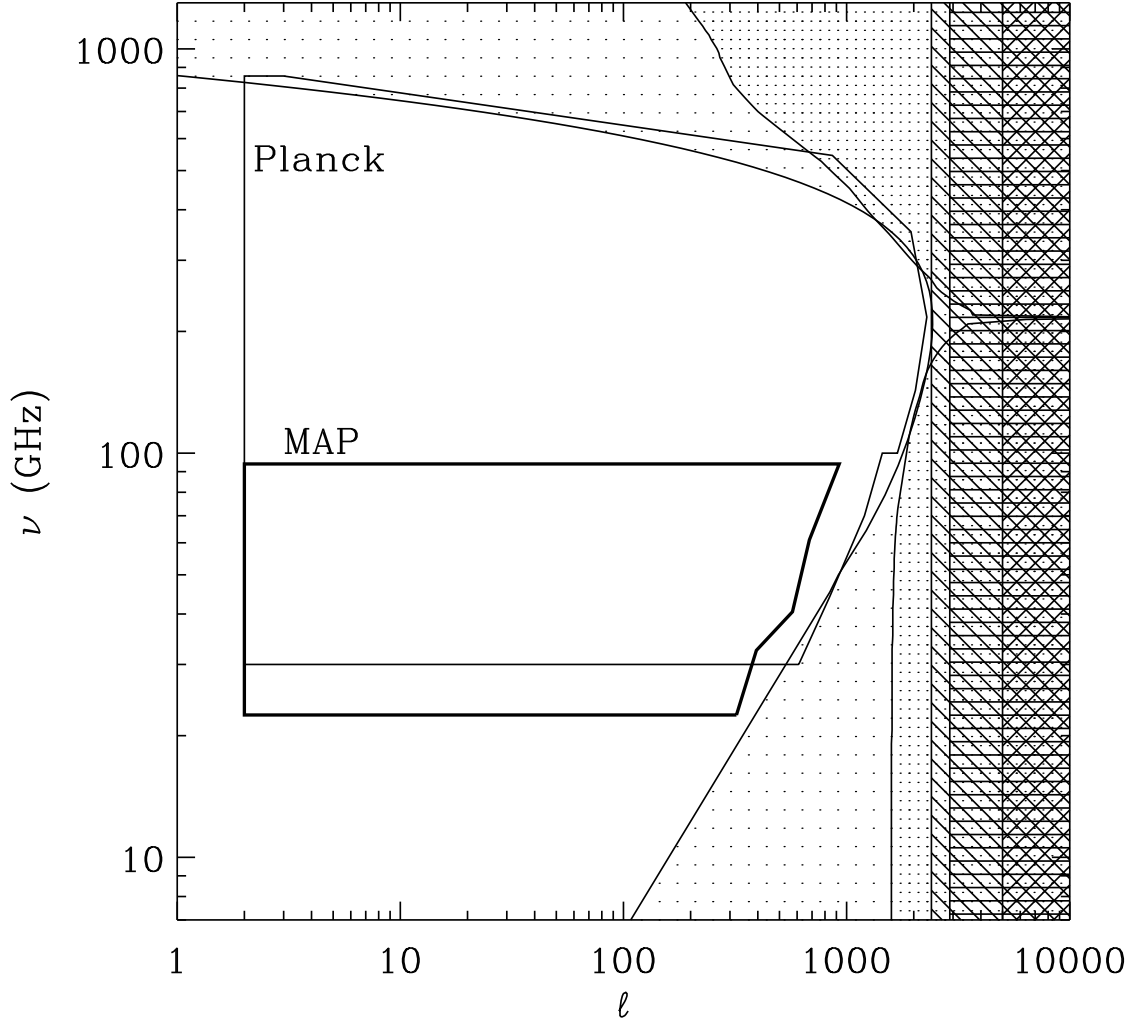}
\caption{Summary of the importance of extragalactic foregrounds of the
CMB. Each filled area on the multipole-frequency plane corresponds to
regions where the foreground fluctuations exceed those of the CMB. The
sparse and dense dotted regions correspond to discrete sources (with
$S<1$ Jy) and the Sunyaev-Zel'dovich effect (with cluster cores),
respectively. The horizontal, descending and ascending hashed regions
correspond to the Ostriker-Vishniac effect (with $z_{r}=50$),
gravitational lensing, and the integrated Sachs-Wolfe effect (with
$\Omega h = 0.25$), respectively. The areas marked \hbox{MAP} (thick
line) and Planck (thin line) show the regions of sensitivity for each
of the future missions, i.e.\ regions where the CMB fluctuations
exceed the noise for each of the instruments (see text).
\label{fig:lnu}}
\end{figure}

We present our results both in table~\ref{tab:foregrounds} and in
figure~\ref{fig:lnu}. In the table, we list $\Delta T_{\ell}$ and
$\ell_{eq}$ for each of the major extragalactic foregrounds at
$\nu=94$ GHz and $\ell=450$, which corresponds to a FWHM angular scale
of about $\theta \sim .3$ deg. These values were chosen to be relevant
to the \hbox{MAP} W-band ($\nu \simeq 94$ GHz and $\theta_{\rm beam}
\simeq 0\fdg21$, see \S\ref{wchan_param}). We also indicate whether
each foreground component has a thermal spectrum.

\placetable{tab:foregrounds}

Figure~\ref{fig:lnu} summarizes the importance of each of the
extragalactic foregrounds in the multipole-frequency plane. It should
be compared to the analogous plot for galactic foregrounds (and
discrete sources) shown in \markcite{teg96}Tegmark \& Efstathiou
(1996; see also \markcite{teg97}Tegmark 1997 for an updated
version). These figures show regions on the $\ell$-$\nu$ plane in
which the foreground fluctuations exceed the CMB fluctuations, i.e.\
in which $C_{\ell}^{\rm foreground} > C_{\ell}^{\rm CMB}$.  As a
reference for $C_{\ell}^{\rm CMB}$, we have used the CDM model shown
in figure~\ref{fig:cl_disc}. Also shown in figure~\ref{fig:lnu} is the
region in which \hbox{MAP} is sensitive, i.e.\ in which $\Delta
C_{\ell}^{\rm noise} < C_{\ell}^{\rm CMB}$, where $\Delta
C_{\ell}^{\rm noise}$ is the {\it rms} uncertainty for \hbox{MAP} (see
Eq.~[\ref{eq:delta_cl}]). A similar region is shown for the current
design parameters of Planck surveyor (\markcite{ber96}Bersanelli et
al.\ 1996) . Note that this figure is only intended to illustrate the
domains of importance of the different foregrounds qualitatively.

\placefigure{fig:lnu}

In the following, we briefly describe each extragalactic foreground
and comment on its respective entries in table~\ref{tab:foregrounds}
and figure~\ref{fig:lnu}.

\subsection{Discrete Sources}
Discrete sources produce positive, point-like, non-thermal, and
possibly time-variable fluctuations. While not much is known about
discrete source counts around $\nu \sim 100$ GHz, several models have
been constructed by interpolating between radio and IR observations
(\markcite{tof98}Toffolatti et al.\ 1998; \markcite{gaw97}Gawiser \&
Smoot 1997; \markcite{gaw98}Gawiser et al. 1998;
\markcite{sok98}Sokasian et al. 1998). Here, we adopt the model of
Toffolatti et al.\ and consider the two flux limits $S <1$ and 0.1 Jy
for the source removal in table~\ref{tab:foregrounds}.  The sparsely
dotted region figure~\ref{fig:lnu} shows the discrete source region
for $S <1$ Jy.  In the context of CMB experiments, the Poisson shot
noise dominates over clustering for discrete sources (see Toffolatti
et al.\ ). As a result, the discrete source power spectrum,
$C_{\ell}^{\rm discrete}$, is essentially independent of
$\ell$. Discrete sources are discussed in more details in
\S\ref{discrete}.

\subsection{Thermal Sunyaev-Zel'dovich Effect}
\label{foregrounds_sz}
The hot gas in clusters and superclusters of galaxies affect the
spectrum of the CMB through inverse Compton scattering. This effect,
known as the Sunyaev-Zel'dovich (SZ) effect (for reviews see
\markcite{sun80} Sunyaev \& Zel'dovich 1980; \markcite{rep95}Rephaeli
1995), results from both the thermal and bulk motion of the gas. We
first consider the thermal SZ effect, which typically has a larger
amplitude and has a non-thermal spectrum (see the \S\ref{OV} below for
a discussion of the kinetic SZ effect). The CMB fluctuations produced
by the thermal SZ effect have been studied using the Press-Schechter
formalism (see \markcite{bar97}Bartlett 1997 for a review), and on
large scales using numerical simulations (\markcite{cen92}Cen \&
Ostriker 1992; \markcite{sca93}Scaramella, Cen, \& Ostriker 1993) and
semi-analytical methods (\markcite{per95}Persi et al.\ 1995). For our
purposes, we consider the SZ power spectrum, $C_{\ell}^{\rm SZ}$,
calculated by Persi et al.\ (see their figure 5). In
table~\ref{tab:foregrounds}, we consider their calculation both with
and without bright cluster removal. In figure~\ref{fig:lnu}, only the
spectrum without cluster removal is shown. A more complete
discussion of the SZ effect is given in \S\ref{SZ}.

\subsection{Ostriker-Vishniac Effect}
\label{OV}
In addition to the thermal SZ effect described above, the hot
intergalactic medium can produce thermal CMB fluctuations as a result
of its bulk motion. While this effect essentially cancels to first
order, the second order term in perturbation theory, the
Ostriker-Vishniac (OV) effect (\markcite{ost86}Ostriker \& Vishniac
1986; \markcite{vis87}Vishniac 1987), can be significant on small
angular scales. The power spectrum of the OV effect depends on the
ionization history of the universe, and has been calculated by
\markcite{huw96}Hu \& White (1996), and \markcite{jaf98}Jaffe \&
Kamionkowski (1998; see also \markcite{per95}Persi et al.\ 1995). We
use the results of Hu \& White (see their figure 5) who assumed that
the universe was fully reionized beyond a redshift $z_{r}$.  In
table~\ref{tab:foregrounds}, we consider the two values $z_{r}=10$ and
50, while in figure~\ref{fig:lnu}, we only plot the region
corresponding to $z_{r}=50$. For consistency, we still use the
standard CDM power spectrum as a reference, even though the primordial
power spectrum would be damped in the event of early
reionization. (Using the damped primordial spectrum makes, at any
rate, only small corrections to both table~\ref{tab:foregrounds} and
figure~\ref{fig:lnu}.)

\subsection{Integrated Sachs-Wolfe Effect}
The Integrated Sachs-Wolfe Effect (ISW) describes thermal CMB
fluctuations produced by time variations of the gravitational
potential along the photon path (\markcite{sac98}Sachs \& Wolfe
1967). Linear density perturbations produce non-zero ISW fluctuations
in $\Omega_m \neq 1$ universe only. Non-linear perturbations produce
fluctuations for any geometry, an effect often called the Rees-Sciama
effect (\markcite{ree68}Rees \& Sciama 1968). \markcite{tul95}Tuluie
\& Laguna (1995) have shown that anisotropies due to intrinsic changes
in the gravitational potentials of the inhomogeneities and
anisotropies generated by the bulk motion of the structures
across the sky generate CMB anisotropies in the range of
$10^{-7} \lesssim \frac{\Delta T}{T} \lesssim 10^{-6}$ on scales of about
$1^{\circ}$ (see also \markcite{tul96} Tuluie et al. 1996). The power
spectrum of the ISW effect in a CDM universe was computed by
\markcite{sel96a}Seljak (1996a; see also references therein). In
table~\ref{tab:foregrounds}, we consider values of the density
parameter, namely $\Omega h=0.25$ and 0.5. In figure~\ref{fig:lnu},
only the $\Omega h=0.25$ case is shown. As above, the standard CDM
($\Omega =1$, $h=0.5$) spectrum is still used as a reference.

\subsection{Gravitational Lensing}
Gravitational lensing is produced by spatial perturbations in the
gravitational potential along the line of sight (see
\markcite{sch92}Schneider, Ehlers, \& Falco 1992;
\markcite{nar96}Narayan \& Bartelmann 1996). This effect does not
directly generate CMB fluctuations, but modifies existing background
fluctuations. The effect of lensing on the CMB power spectrum was
calculated by \markcite{sel96b}Seljak (1996b) and
\markcite{met97}Metcalf \& Silk (1997). Recently,
\markcite{zal98}Zaldarriaga \& Seljak (1998) included the lensing
effect in their CMB spectrum code (CMBFAST; \markcite{seljz96} Seljak
\& Zaldarriaga 1996). We use this code to compute the absolute lensing
correction $|\Delta C_{\ell}^{\rm lens}|$ to the standard CDM spectrum
of figure~\ref{fig:cl_disc}, including nonlinear evolution. The
results are shown in table~\ref{tab:foregrounds} and
figure~\ref{fig:lnu}.

\subsection{Other Extragalactic Foregrounds}
In addition to the effects discussed above, other extragalactic
foregrounds can cause secondary anisotropies.  For instance, patchy
reionization produced by the first generation of stars or quasars can
cause second order CMB fluctuations through the doppler effect
(\markcite{agh96a,agha96b}Aghanim et al. 1996a,b;
\markcite{gru98}Gruzinov \& Hu 1998; \markcite{kno98}Knox, Scoccimaro,
\& Dodelson 1998; \markcite{pee98}Peebles \& Juzkiewicz 1998).
Calculations of the spectrum of this effect are highly uncertain, but
show that the resulting CMB fluctuations could be of the order of 1
$\mu$K on 10 arcminute scales, for extreme patchiness. More likely
patchiness parameters make the effect negligible on these scales, but
potentially important on arcminute scales. Another potential
extragalactic foreground is that produced by the kinetic SZ effect
from Ly$_{\alpha}$ absorption systems, as was recently proposed by
\markcite{loe96}Loeb (1996). The resulting CMB fluctuations are of the
order of a few $\mu$K on arcminute scales, and about one order of
magnitude lower on 10 arcminute scales.  Because of the uncertainties
in the models for these two foregrounds and because they are small on
10 arcminute scales, we do not include them in
table~\ref{tab:foregrounds} and figure~\ref{fig:lnu}.

\subsection{Comparison}
An inspection of table~\ref{tab:foregrounds} shows that, for
\hbox{MAP}, the power spectra of the largest extragalactic foregrounds
considered are a factor of 5 below the primordial CDM spectrum. As can
be seen in figure~\ref{fig:lnu}, the dominant foregrounds for
\hbox{MAP} are discrete sources, the thermal SZ effect and
gravitational lensing. Note that for Planck surveyor, these three
effects produce fluctuations which are close to the sensitivity of the
instrument. The spectra of the OV and ISW effects will produce
fluctuations of the order of $1 \mu$K for \hbox{MAP}, and are thus
less important.  The effect of gravitational lensing is now
incorporated in CMB codes such as CMBFAST, and can thus be taken into
account in the estimation of cosmological parameters. The other two
dominant extragalactic contributions, discrete sources and the thermal
SZ effect, must also be accounted for, but are more difficult to
model. We focus on these two foregrounds in the following
sections. Note that, on large angular scales, extragalactic
foregrounds produce relatively small fluctuations, and are thus
not detectable in the COBE maps (\markcite{bou93}Boughn \& Jahoda 1993;
\markcite{ben93}Bennett et al.\ 1993; \markcite{ban96}Banday et al.\
1996; \markcite{kne97}Kneissl et al.\ 1997)

\section{Discrete Sources}
\label{discrete}
In this section, we use the existing literature to estimate the
discrete source contribution to the 94 GHz \hbox{MAP} channel. Since
discrete sources generally have spectra that differ from that of the
CMB, this single-channel analysis should be viewed as being
conservative. Note, however, that for the sizable fraction of radio
sources which have a flat spectrum (see \markcite{sok98}Sokasian et
al. 1998), a multi-channel analysis will not dramatically improve the
source subtraction. Indeed, the Rayleigh-Jeans (RJ) temperature
fluctuations produced by such sources (with fluxes $S_{\nu} \propto
\nu^{0}$) depend on frequency as $\Delta T \propto S_{\nu} \theta_{\rm
beam}^{-2} \nu^{-2} \propto \nu^{0}$ since $\theta_{\rm beam} \propto
\nu^{-1}$ approximately for \hbox{MAP}, and are thus indistinguishable
from CMB fluctuations. Source subtraction can also be improved by
searching the time-ordered data for time-variability, which is common
for radio sources. It can also be improved by cross-correlating the
CMB maps with existing source catalogs (eg. that by
\markcite{sok98}Sokasian et al. 1998; for searches in the COBE maps
see \markcite{ben93}Bennett et al.\ 1993; \markcite{ban96}Banday et
al.\ 1996; see also \S\ref{sz_stack} below for an application of this
technique to SZ clusters).
\label{flat_spectrum}

\subsection{Source Counts}
\markcite{tof98}Toffolatti et al.\ (1998) have computed the expected
discrete source counts based on radio and IR surveys.  An
extrapolation of the Toffolatti et al.\ counts to 94 GHz yields an
integrated source density of
\begin{equation}
n_{\rm disc}(>5\sigma) \simeq 2.2 \times 10^{-3}
{\rm ~sources~deg}^{-2},
\end{equation}
where the flux threshold is the $5\sigma$ detection threshold for the
\hbox{MAP} 94 GHz channel, $S_{94}(5\sigma) \approx 2$ Jy (see
Eq.~[\ref{eq:s_onesigma}]). The corresponding number of sources
directly detectable by \hbox{MAP} is
\begin{equation}
N_{\rm disc}(>5\sigma) \simeq 46 \left( \frac{f_{\rm sky}}{0.5}
\right) {\rm ~sources},
\label{eq:n_disc}
\end{equation}
where $f_{\rm sky}$ is the fractional sky coverage resulting from the
galactic cut. As a comparison, the expected number of random
$>5\sigma$ fluctuations in $0\fdg3 \times 0\fdg3$ pixels is $N_{\rm
random}(>5\sigma) \simeq 0.27 f_{\rm sky}$. 

The contribution of discrete sources to the microwave sky has also
been investigated by \markcite{gaw97}Gawiser \& Smoot (1997),
\markcite{gaw98}Gawiser et al. (1998), and \markcite{sok98}Sokasian et
al. (1998). In particular, Sokasian et al. have used a catalog of 2200
bright extragalactic sources, over 700 of which have been observed at
90 GHz. They predict $N_{\rm disc}(>5\sigma) \simeq 25$ for the 94 GHz MAP
channel with $f_{\rm sky}=0.5$. This is about a factor of 2 below the
above prediction from Toffolatti et al. This discrepancy is acceptable
given the uncertainties involved in the source surveys and models.

\subsection{Residual Variance}
Unless external catalogs are used, sources with fluxes less than
$S_{94}(5\sigma)$ will contaminate the CMB maps. Ignoring source
clustering (see \markcite{tof98}Toffolatti et al.\ 1998), the CMB
fluctuations from discrete sources are dominated by Poisson shot noise
which is independent of $\ell$.  (For the 94 GHz channel, the
clustering variance is about 4 times lower than the Poisson variance,
and is ignored in this analysis; see Toffolatti et al.\ ). A mild
extrapolation of the source power spectrum of Toffolatti et al.\ yields
\begin{equation}
C_{\ell}^{\rm disc}(94 {\rm GHz}, <5\sigma) \simeq 1.71 \times
10^{-16},
\label{eq:cl_disc}
\end{equation}
for all $\ell$. Euclidean counts ($N(>S) \propto S^{-3/2}$) were
assumed to extrapolate the 94 GHz flux limit to $S_{94}(5\sigma)$.
The resulting temperature {\it rms} fluctuations for the \hbox{MAP} 94
GHz channel is (See Eq.~[\ref{eq:deltaT_cmb}] with $F_{\ell} \equiv
1$)
\begin{equation}
\Delta T_{\rm disc} \simeq 6.5 \mu {\rm K}.
\end{equation}
Note that this figure includes contribution from all the multipoles,
while the related entries in table~\ref{tab:foregrounds} refer only to
a $\Delta\ln\ell = 1$ interval around $\ell=450$.  The discrete source
power spectrum is shown as the dashed line in
figure~\ref{fig:cl_disc}.  The discrete source fluctuations are well
below the CMB fluctuations for all relevant $\ell$'s. However, they
are comparable to the 94 GHz channel noise for $\ell \sim
700$. Discrete sources must therefore be carefully accounted for when
estimating cosmological parameters with \hbox{MAP}. This can be
achieved by calibrating the discrete source power spectrum
(Eq.~[\ref{eq:cl_disc}]) by counting the number of detectable sources
in the map (Eq.~[\ref{eq:n_disc}]).

\subsection{Residual Skewness}
The residual discrete sources will not only contribute to the CMB
variance but also to its skewness. To estimate the resulting skewness,
let us consider the discrete source statistics in $N_{\rm pix}$ pixels
of solid angle $\Omega_{\rm pix}$. The mean number of sources brighter
than the flux detection threshold $S_{d}$ in such a pixel is
$N(>S_{d}) = \Omega_{\rm pix} \int_{S_{d}}^{\infty} dS \frac{dn}{dS}$,
where $\frac{dn}{dS}$ is the differential source count per unit solid
angle. The statistics of the unresolved background due to all sources
with fluxes below $S_{d}$ can be quantified using the moments of the
differential counts defined as $Q_{\alpha} \equiv \Omega_{\rm pix}
\int_{0}^{S_{d}} dS S^{\alpha} \frac{dn}{dS}$.  The mean unresolved
intensity in a pixel is then $<I> = Q_{1}$.  The intensity variance
and skewness are $\sigma^{2}_{\rm disc} \equiv \langle \left( I-
\langle I \rangle \right)^{2} \rangle = Q_{2}$ and $\eta_{\rm disc}
\equiv \langle \left( I - \langle I \rangle \right)^{3} \rangle =
Q_{3}$, respectively. To derive the last expressions, we assumed that
the source counts obey Poisson statistics, thereby ignoring source
clustering. For the \hbox{MAP} detection threshold ($S_{d}=S_{94}(5
\sigma) \simeq 2$ Jy), the former integrals are dominated by counts
close to $S_{d}$ which are nearly Euclidean. We therefore approximate
$\frac{dn}{dS} \propto S^{-5/2}$. In this case, it is easy to show
that $\sigma_{\rm disc}^{2} = 3 N(>S_{d}) S_{d}^{2}$ and
$\eta_{\rm disc} = N(>S_{d}) S_{d}^{3} = \frac{1}{3}
\sigma_{\rm disc}^{2} S_{d}$.

Let us then determine whether discrete sources can produce a
detectable skewness. For this purpose, let us assume that the
intrinsic CMB fluctuations (along with instrumental noise) are
gaussian distributed. In this case, CMB fluctuations have vanishing
skewness, i.e. $\eta_{\rm CMB} = 0$. The $1\sigma$ uncertainty in
measuring this zero skewness is $\sigma[ \eta_{\rm CMB} ] = \sqrt{
\frac{6}{N_{pix}} } \sigma_{\rm CMB}^{3}$, where $\sigma^{2}_{\rm CMB}$ is
the variance of the CMB fluctuations.  As a result, the
signal-to-noise ratio for detecting the discrete source skewness in
the maps is SNR$_{\eta} \equiv \eta_{\rm disc} \sigma^{-1}[
\eta_{\rm CMB} ]$. Combining this with the above Euclidean
approximations, we get SNR$_{\eta} \simeq (6 N_{\rm pix})^{-\frac{1}{2}}
{\rm SNR}_{d}^{3} N_{\rm disc}(>S_{d})$, where SNR$_{d} \equiv
S_{d} \sigma_{\rm CMB}^{-1}$ is the signal-to-noise ratio
for the detection, and $N_{\rm disc}(>S_{d}) = N_{\rm pix} N(>S_{d})$
is the total number of detected sources in the map. The number
of pixels $N_{\rm pix}$ is related to the fraction of the 
sky covered $f_{\rm sky}$ by $N_{\rm pix} = 4 \pi f_{\rm sky}
\Omega_{\rm pix}^{-1}$. For the \hbox{MAP} 94 GHz channel with $0\fdg3
\times 0\fdg3$ pixels, we get
\begin{equation}
{\rm SNR}_{\eta} \simeq
4.9 \left( \frac{f_{\rm sky}}{0.5} \right)^{\frac{1}{2}}
\left( \frac{{\rm SNR}_{d}}{5} \right)^{3}
\left( \frac{n(>S_{d})}{2.2\times 10^{-3} {\rm deg}^{-2}}  \right).
\end{equation}
We therefore expect the skewness from discrete sources to be
detectable at the $\sim 5 \sigma$ level. Care must be taken that this
skewness is not mistaken for the intrinsic CMB skewness.
  
\section{Thermal Sunyaev-Zel'dovich Effect}
\label{SZ}
As we discussed in \S\ref{foregrounds_sz}, the thermal SZ effect
describes the inverse Compton scattering of CMB photons with the hot
gas in clusters and superclusters of galaxies. The resulting change
in the (thermodynamic) CMB temperature observed at frequency $\nu$ is given by
\begin{equation}
\label{eq:deltat_sz}
\frac{\Delta T_{SZ}}{T_{0}} = y j(x),
\end{equation}
where $T_{0}$ is the unperturbed CMB temperature, $y$ is the
comptonization parameter, $x$ is a dimensionless parameter defined as
$x \equiv \frac{h \nu}{k T_{0}}$, and $j(x)$ is a spectral function. The
Compton parameter is given by
\begin{equation}
\label{eq:def_y}
y=\int dl \frac{k T_{e}}{m_{e}c^{2}} \sigma_{T} n_{e},
\end{equation}
where $T_{e}$ is the electron temperature, $n_{e}$ is the electron
density, $\sigma_{T}$ is the Thompson cross-section, and the integral
is over the line-of-sight distance. In the nonrelativistic regime
($kT_{e} \ll m_{e} c^{2}$), the spectral function is
\begin{equation}
\label{eq:jx}
j(x) = \frac{x \left( e^{x}+1 \right)}{e^{x}-1} - 4
\end{equation}
which is zero at $x_{0}\simeq 3.83$, corresponding to $\nu_{0} \simeq
217$ GHz for $T_{0}=2.725$ K. The SZ temperature shift $\Delta T_{SZ}$
is negative (positive) for observation frequencies $\nu$ below (above)
$\nu_{0}$. In the RJ limit ($x \ll 1$), the spectral
function becomes $j(x \ll 1) \simeq -2$. For the \hbox{MAP} W-channel
$(\nu\simeq 94$ GHz), $x=x_{94} \simeq 1.65$ and $j(x_{94}) \equiv
j_{94} \simeq -1.56$.

In the following, we study the impact of the SZ effect on \hbox{MAP}.  In
\S\ref{SZ_prominent}, we estimate the SZ decrement for prominent
clusters of galaxies. In \S\ref{SZ_clusters}, we consider a
cosmological population of cluster via both the Press-Schechter
formalism and existing X-ray cluster catalogs. Finally, we study
the SZ effect from hot gas on supercluster scales in
\S\ref{SZ_superclusters}.

\subsection{Prominent SZ sources}
\label{SZ_prominent}
We begin by surveying the most prominent SZ sources in the sky and
establishing their relevance for \hbox{MAP}. Note that a search for the SZ
effect of nearby clusters of galaxies in the COBE-DMR maps has only
led to upper limits (\markcite{kog94}Kogut et al.\ 1994).

For this purpose, we will make use of the isothermal $\beta$-model
which is commonly used to fit cluster profiles. In this model, the
3-dimensional electron density $n_{e}(r)$ is taken to be spherical
with a profile given by
\begin{equation}
n_{e}(r) = n_{0} \left[ 1+ \left( \frac{r}{r_{c}} \right)^{2}
\right]^{-\frac{3}{2}\beta},
\end{equation}
where $n_{0}$ is the central electron density, $r_{c}$ is the
core radius, and $\beta$ is the slope parameter. For an isothermal
cluster with temperature $T_{e}$, the resulting angular SZ profile
(Eq.~[\ref{eq:deltat_sz}]) is
\begin{equation}
\label{eq:dtsz_theta}
\Delta T_{SZ}(\theta) = \Delta T_{SZ}(0) \left[ 1+ \left(
\frac{\theta}{\theta_{c}} \right)^{2} \right]^{-\frac{3}{2}\beta +
\frac{1}{2}},
\end{equation}
where $\theta_{c} \equiv r_{c}/D_{A}$ is the projected core radius and
$D_{A}$ is the angular diameter radius. The central SZ decrement
$\Delta T_{SZ}(0)$ (Eqs.~[\ref{eq:deltat_sz}] \& [\ref{eq:def_y}]) is
related to the central electron density $n_{0}$ as
\begin{equation}
\label{eq:dtsz_0}
\Delta T_{SZ}(0) \simeq -38.8 \mu\mbox{\rm K} 
\left( \frac{n_{0}}{10^{-3} \mbox{\rm cm}^{-3}} \right)
\left( \frac{k T_{e}}{1 \mbox{\rm keV}} \right)
\left( \frac{r_{c}}{1 \mbox{\rm Mpc}} \right)
\left( \frac{j(x)}{-2} \right)
\frac{\Gamma(3 \beta /2 - 1/2)}{\Gamma(3 \beta /2)},
\end{equation}
where $j(x)$ is the spectral function defined in Eq.~(\ref{eq:jx}),
and $\Gamma$ is the gamma function. In practice, the central electron
density can be determined from observations of the X-ray intensity
profile which scales as $i_{x}(\theta) \propto n_{0}^{2}
\left[ 1 + \left(
\frac{\theta}{\theta_{c}} \right)^{2} \right]^{-3 \beta + \frac{1}{2}}$.

\begin{figure}
\plotone{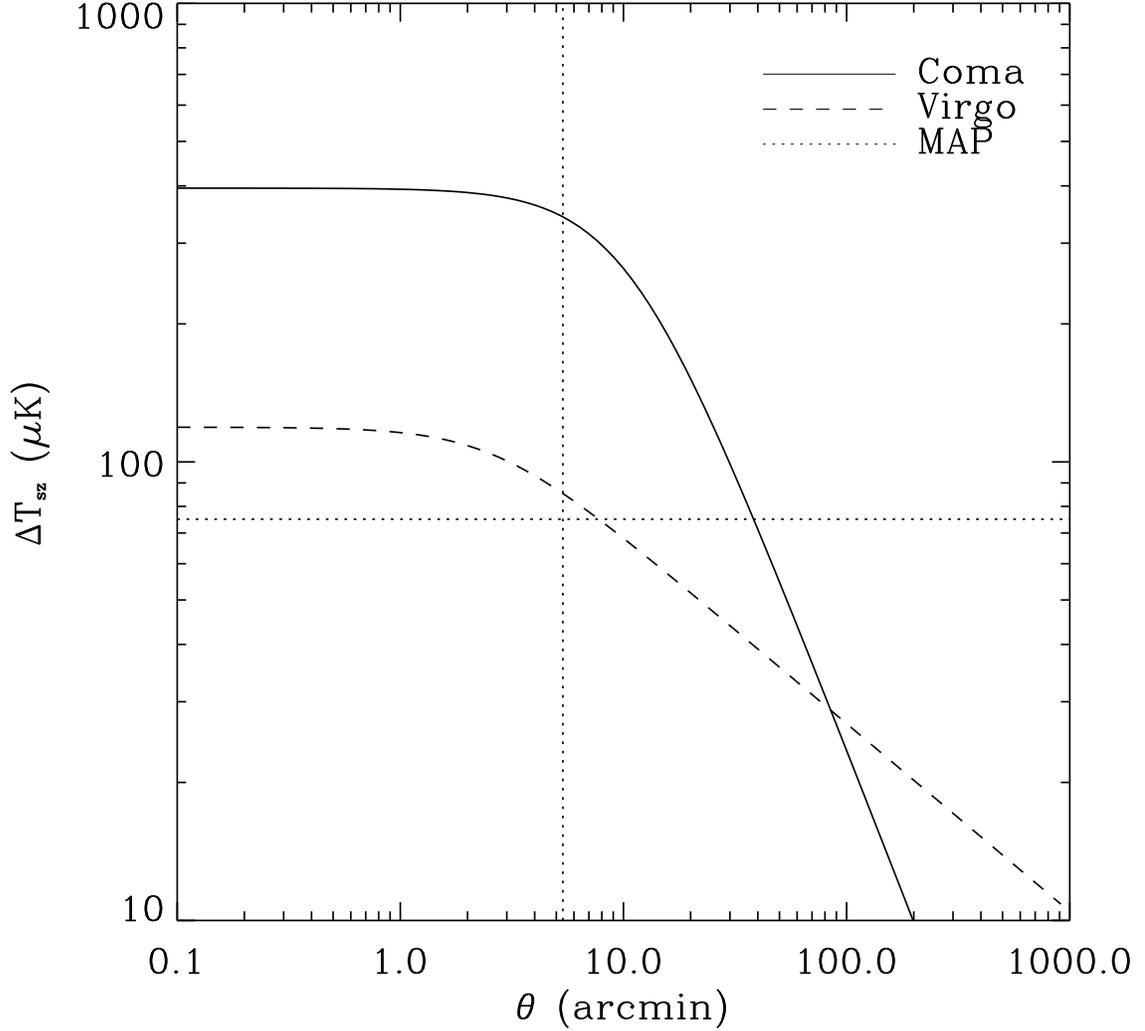}
\caption{Sunyaev-Zel'dovich temperature profiles expected for the Coma
and Virgo clusters at $\nu=94$ GHz. Both clusters will produce
temperature decrements, but the absolute temperature shifts are
plotted. The horizontal dotted lines show the $1\sigma$ detection
threshold and $1\sigma$ beam radius for the 94 GHz \hbox{MAP} channel. Virgo
is more extended and should be close to the detection threshold, while
Coma should be clearly detectable and marginally resolvable.
\label{fig:comavirgo}}
\end{figure}

\begin{itemize}
\item The Local Group: The hot gas in the Local Group (LG) surrounds
the Milky Way and can thus, in principle, contribute to the CMB
quadrupole (\markcite{Sut96}Suto et al.\ 1996). In general, one expects
the CMB and LG quadrupoles to contribute to the total quadrupole
$Q_{\rm tot}$ as $Q_{\rm tot}^{2}=Q_{\rm CMB}^2 + Q_{LG}^{2}$, where
$Q^{2} \equiv 5 C_{2} T_{0}^{2}/ 4 \pi$. However,
\markcite{pil96}Pildis \& McGaugh (1996) showed that the LG quadrupole
is of the order of $Q_{\rm LG} \sim 0.1 \mu$K for $\beta$-models of the
LG which are consistent with observations of poor, spiral-dominated
groups. This is well below the quadrupole measured by COBE, $Q_{\rm
COBE} = 10.0^{+3.8}_{-2.8} \mu$K ($1\sigma$; \markcite{ben96}Bennett
et al.\ 1996), and the cosmic variance uncertainty, $\Delta Q \simeq 3.2
\mu$K (see Eq.~[\ref{eq:delta_cl}]). Thus, unless the LG is a very
peculiar group, its SZ effect on the CMB is negligible.

\item Virgo: Virgo is at a distance of about 18 Mpc and is thus the
nearest cluster. X-ray observations of Virgo
(\markcite{boh94}B\"{o}hringer et al.\ 1994; \markcite{nul95}Nulsen \&
B\"{o}hringer 1995) show highly irregular emission with a temperature
of $kT_{e} \simeq 2.4$ keV.  A fit to the X-ray profile derived from
the ROSAT All-Sky Survey (\markcite{boh98}B\"{o}hringer 1998) yields
the following $\beta$-model parameters: $\theta_{c} \simeq 2\farcm65$,
$\beta \simeq 0.47$, and $n_{0} \simeq 0.029$ cm$^{-3}$. The resulting
central SZ temperature at $\nu=94$ GHz is $\Delta T_{\rm SZ}(0) \simeq
- 113 \mu$K (Eq.~[\ref{eq:dtsz_0}]). The SZ profile derived for these
parameters (Eq.~[\ref{eq:dtsz_theta}]) is shown in
figure~\ref{fig:comavirgo}. Also shown are the $1\sigma$ detection
threshold (Eq.~[\ref{eq:deltaT_noise}]) and $1\sigma$ beam radius for
the 94 GHz \hbox{MAP} channel. Virgo is thus rather extended, but has
a central SZ decrement which is only $1.5\sigma$ above the \hbox{MAP}
detection threshold.

\item Coma: Coma is at $z\simeq 0.02$ and is the nearest massive
cluster. Fits of a $\beta$-model to X-ray observations
(\markcite{hug88}Hughes, Gorenstein, \& Fabricant 1988;
\markcite{bri92}Briel, Henry, \& B\"{o}hringer 1992) yields: $kT_{x}
\simeq 9.1$ keV, $\theta_{c} \simeq 10\farcm5$, and $\beta \simeq
0.75$. The SZ effect for Coma was detected by \markcite{her95}Herbig
et al.\ (1995) who found a RJ central decrement of $\Delta T_{\rm
SZ}(0)=-505\pm92$ $\mu$K. The resulting 94 GHz SZ profile
(Eq.~[\ref{eq:dtsz_theta}]) is shown in
figure~\ref{fig:comavirgo}. Coma will be clearly detected ($>5\sigma$
in the central pixel) and marginally resolved by \hbox{MAP}.

\item Unresolved Bright Clusters: In addition, about 10 clusters will
be detectable by \hbox{MAP} at the $>5\sigma$
level. Table~\ref{tab:clusters} shows the 12 clusters which are
expected to yield the largest SZ flux in the XBACs catalog (see
\S\ref{empirical}). A study of the SZ effect produced by a
cosmological population of clusters is the object of the next section.

\end{itemize}

\placefigure{fig:comavirgo}

\subsection{Cosmological Cluster Population} 
\label{SZ_clusters}
In addition to the above prominent clusters, the population of
clusters distributed throughout the universe will produce SZ
fluctuations in the CMB. We estimate the SZ cluster counts for \hbox{MAP} by
considering both predictions from the Press-Schechter formalism and
existing X-ray cluster catalogs.

\subsubsection{Press-Schechter Predictions}
The Press-Schechter formalism (PS; Press \& Schechter
1974\markcite{pre74}) can be used to predict the SZ cluster counts in
a given cosmology (\markcite{del95}De Luca, D\'{e}sert, \& Puget 1995;
\markcite{bar96}Barbosa et al.\ 1996; \markcite{col97}Colafrancesco
et al.\ 1997; see also \markcite{agh97}Aghanim et al.\ 1997 and the
review by \markcite{bar97}Bartlett 1997). For our purposes, we use the
PS predictions of De Luca et al.\ (1995)\markcite{del95} who
considered a CDM model with $n=1$, $b=1.7$, $\Omega_{0}=1$, and
$h=0.5$. Their normalization was chosen to match the observed X-ray and
optical cluster counts. For fluxes above $S_{94}\simeq 2$ mJy, the
integrated counts are nearly Euclidean and given by
\begin{equation}
n(>S_{94}) \simeq 0.043 
  \left( \frac{S_{94}}{100{\rm mJy}} \right)^{-\frac{3}{2}}
{\rm deg}^{-2},
\label{eq:ns_ps}
\end{equation}
where fluxes have been converted to $S_{94}\equiv S(94{\rm GHz})$.
The differential $dn/dS$ and integrated $n(>S)$ counts are
shown in Figure~\ref{fig:counts}.

\subsubsection{Empirical predictions}
\label{empirical}
Cluster counts can also be empirically predicted by considering
existing cluster catalogs. For this purpose, we utilize the XBACs
catalog (\markcite{ebe95}Ebeling et al.\ 1996) which consists of a
nearly complete, X-ray flux limited, sample of Abell-type clusters
detected in the ROSAT All Sky Survey. The survey consists of 242
optically selected clusters with $z<0.2$, $|b|> 20^{\circ}$, and
$S_{x}(0.1-2.4 {\rm keV})< 5\times 10^{-12}$ erg s$^{-1}$ cm$^{-2}$.
When available, the future BCS cluster catalog
(see \markcite{ebe97}Ebeling et al.\ 1997) will be X-ray selected and will
thus be even more suited to the present study.

The SZ fluxes of clusters can be estimated from their X-ray properties
using the virial theorem. Indeed, the X-ray temperature of a
virialized cluster is related to its mass by $T_{e} \propto
M^{\frac{2}{3}}$.  Such a relation was verified to hold well in the
numerical simulations of \markcite{evr96}Evrard, Metzler, \& Navarro
(1996) who found the proportionality constant to be
\begin{equation}
kT_{e} \simeq 1.27 \left( \frac{M}{10^{14} M_{\odot}}
\right)^{\frac{2}{3}} {\rm keV}.
\end{equation}
Here, $M$ is the mass within the virialized region, which is
approximated to the region corresponding to a density contrast of
$\delta=500$. The dispersion between the above mass estimates and that
from numerical simulations was found by Evrard et al.\ to have a
standard deviation of less than 15\%.

The total flux $S_{\nu}$ of a cluster observed at frequency $\nu$ is
related to the angular integral of the thermodynamic temperature shift
$\Delta T_{\nu}$ by
\begin{equation}
S_{\nu} = \frac{2 k^{3} T_{0}^{2}}{h^{2} c^{2}} q(x) \int d\Omega
~\Delta T_{\nu}(\mbox{\bf $\theta$}),
\end{equation}
where $q(x) \equiv x^4 / (2 \sinh \frac{x}{2})^{2}$ is a spectral function, and
$x \equiv h \nu / k T_{0}$ as before. For the \hbox{MAP} W-channel ($\nu =
94$ GHz, $x=x_{94} \simeq 1.65$), $q(x_{94}) \equiv q_{94} \simeq
2.18$. By inserting equations~(\ref{eq:deltat_sz}) and
(\ref{eq:def_y}) into the above expression, it is easy to see that the
total SZ flux of a virialized cluster scales as $S_{\rm SZ} \propto
f_{\rm gas} M T_{e} \propto f_{\rm gas} T_{e}^{\frac{5}{2}}$. where
$f_{\rm gas} \equiv M_{\rm gas}/M$ is the gas mass
fraction. Specifically,
\begin{equation}
\label{eq:s_sz}
S_{94} \simeq 11.44 \left( \frac{300 {\rm Mpc}}{D(z)} \right)^{2}
\left( \frac{f_{\rm gas}}{.11} \right) \left( \frac{kT_{e}}{1 {\rm keV}}
\right)^{\frac{5}{2}} {\rm mJy},
\label{eq:s94_te}
\end{equation}
where $D(z)$ is the angular-diameter distance to the cluster. The
central value for $f_{\rm gas}$ was chosen to be the mean of the
observed gas fractions in the cluster sample of \markcite{whi95}White
\& Fabian (1995). While the SZ surface brightness (or temperature) is
independent of redshift, the total SZ flux scales as $S_{\rm SZ}
\propto D^{-2}$ since the angular size of the cluster scales as
$\theta \propto D^{-1}$.

\begin{figure}
\plotone{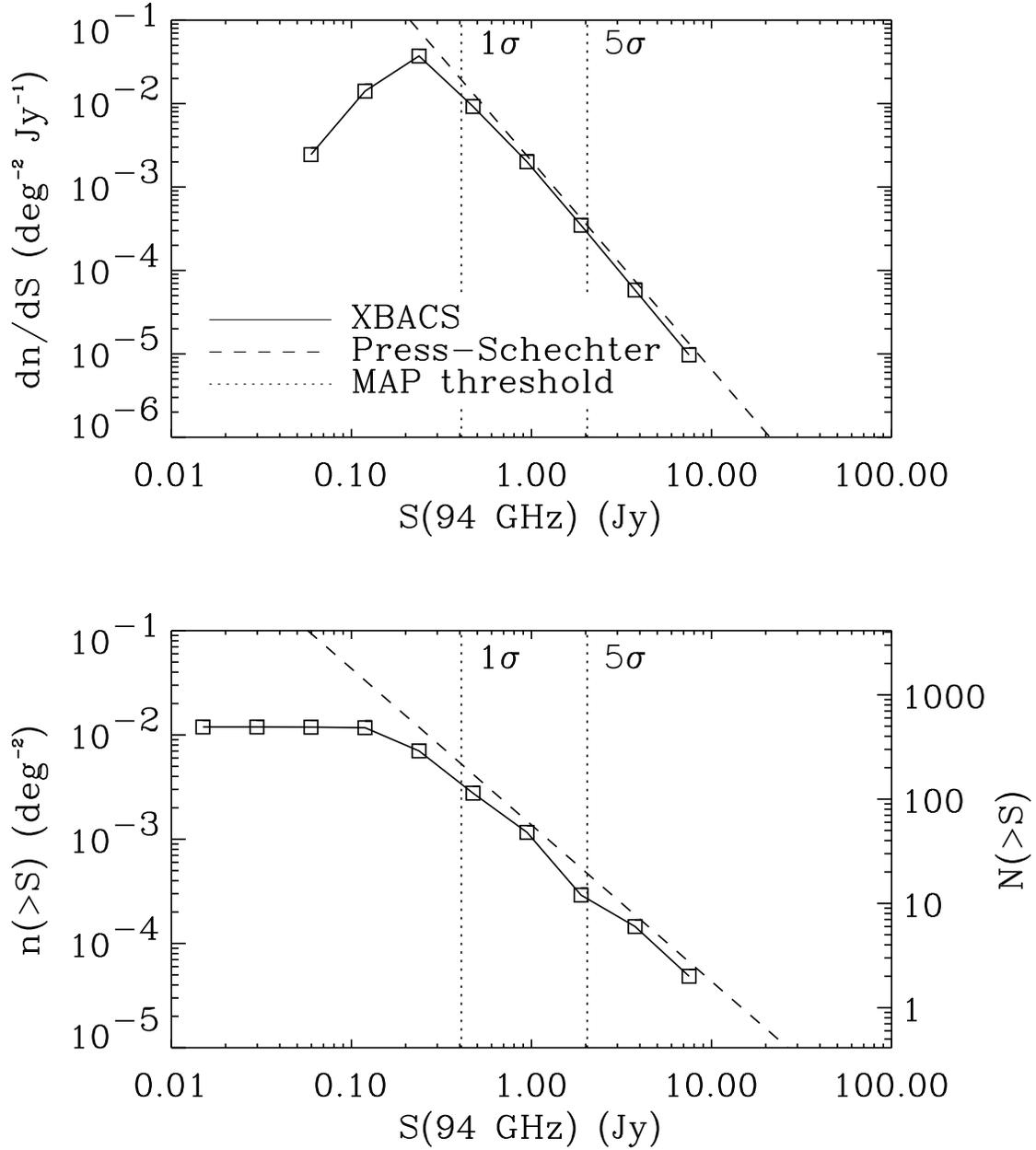}
\caption{Differential and cumulative 94 GHz SZ cluster counts
expected for the XBACS catalog (Ebeling et al.\ 1996) and for the
Press-Schechter counts of De Luca et al.\ (1995). The right-hand
vertical scale on the bottom panel gives the total number of clusters
on the full sky.  The gas mass fraction and Hubble constant were taken
to be $f_{\rm gas}=0.11$ and $h=0.5$, respectively.  The $1\sigma$ and
$5\sigma$ detection thresholds for the 94 GHz \hbox{MAP} channel are
shown as vertical dotted lines.
\label{fig:counts}}
\end{figure}

The SZ fluxes of clusters in the XBACs catalog can thus be estimated
from the listed values of $T_{e}$ and $z$. The resulting cumulative
and differential SZ counts are shown in figure~\ref{fig:counts}, for
$h=0.5$ and $f_{\rm gas}=0.11$. The XBACS counts are nearly Euclidean
for $S_{94} \gtrsim 100$ mJy. A comparison with the Press-Schechter
counts shows that about 80\% of SZ clusters above this flux are contained
in the XBACs catalog.

\placefigure{fig:counts}

\subsubsection{Detection}
The most obvious method to detect SZ clusters is to try to detect them
one by one. The $1\sigma$ and $5\sigma$ detection thresholds for \hbox{MAP}
(see Eq.~[\ref{eq:s_onesigma}]) are indicated on
Figure~\ref{fig:counts}. According to Press-Schechter predictions
(Eq.~[\ref{eq:ns_ps}]), the number of clusters which \hbox{MAP} will directly
detect at the $5\sigma$ level is
\begin{equation}
N(>5\sigma) \simeq 9.7 \left( \frac{f_{\rm sky}}{0.5} \right)
{\rm ~clusters},
\end {equation}
where $f_{\rm sky}$ is the covered fraction of the sky. Recall that
the $5\sigma$ detection threshold for the \hbox{MAP} 94 GHz channel is
$S_{94}(5\sigma) \approx 2$ Jy (see Eq.~\ref{eq:s_onesigma}).  Out
of these detectable clusters, 8 clusters are expected to be in the
XBACs catalog. Table~\ref{tab:clusters} lists the 12 brightest SZ
clusters in the XBACs catalog. In this table, SZ fluxes were derived
using equation~(\ref{eq:s94_te}) with $h=0.5$ and $f_{\rm gas}=0.11$.
When possible, more accurate redshifts and X-ray temperatures were
taken from \markcite{dav93}David et al.\ (1993). \hbox{MAP} will thus
allow for a direct measurement of the total (or ``zero-spacing'' in
interferometric parlance) SZ flux of the brightest clusters and for a
check of the relation between their X-ray and SZ properties. In
addition, the counts of these detectable clusters can be used to
calibrate the Press-Schechter SZ counts (Eq.~[\ref{eq:ns_ps}]) and
thus to constrain the residual contribution of SZ clusters to the CMB
power spectrum. Note, however, that some clusters contain bright radio
sources, and will thus require high-resolution maps for an accurate
determination of their SZ flux.

\begin{figure}
\plotone{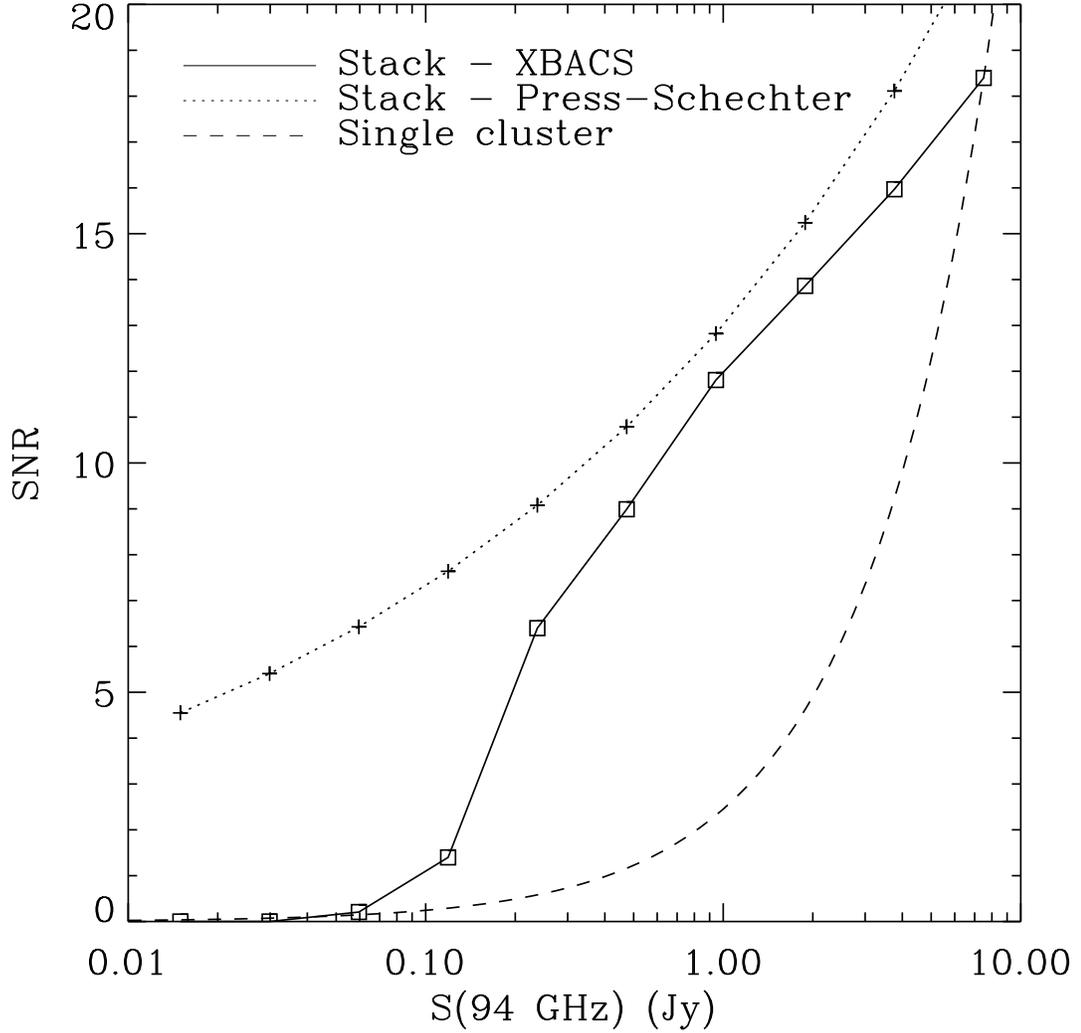}
\caption{Signal-to-noise ratio (SNR) for the detection of
clusters using the stacking technique with the 94 GHz \hbox{MAP} channel. The
solid line corresponds to the XBACs catalog, while the dotted line
corresponds to the Press-Schechter counts of De Luca et al.\ (1995) for
a sky coverage of 80\%. The dashed line shows the SNR for detecting a
single cluster.
\label{fig:stack}}
\end{figure}

Fainter clusters can be studied statistically by performing a
cross-correlation between CMB temperature and cluster positions.
Specifically, one can imagine ``stacking'' regions of the sky centered
on clusters with known positions.  For a stack of $N_{c}$ clusters
with mean SZ flux $\overline{S}_{94}$, the signal-to-noise ratio is
\begin{equation}
{\rm SNR}_{\rm stack} = \left( \frac{\overline{S}_{94}}{S_{94}(1\sigma)}
\right) \sqrt{N_{c}},
\end{equation}
where $S_{94}(1\sigma)$ is the $1\sigma$ detection threshold given in
equation~(\ref{eq:s_onesigma}). Figure~\ref{fig:stack} shows the
resulting ${\rm SNR}_{\rm stack}$ for the cross-correlation of XBACs
clusters with \hbox{MAP}. A significant signal (${\rm SNR}_{\rm stack}
> 5$) is expected for $S_{94} > 200$ mJy, one order of magnitude below
the direct detection threshold ($S_{94}(5\sigma) \simeq 2.04$
Jy). This can be used to test the virialization state of clusters, and
to constrain the mean gas fraction and mass-temperature relation of
clusters.  Note that a similar cross-correlation analysis was
performed with the COBE maps and the ACO cluster catalog but only led
to upper limits (\markcite{ben93}Bennett et al.\ 1993;
\markcite{ban96}Banday et al.\ 1996).
\label{sz_stack}

\placetable{tab:clusters}

\placefigure{fig:stack}

\subsection{Hot Gas on Supercluster Scales} 
\label{SZ_superclusters}
Hot gas on supercluster scales can also produce CMB fluctuations
through the SZ effect. The presence of hot gas on large scales and its
shock heating to about 1 keV is predicted by hydrodynamical numerical
simulations (\markcite{cen92}Cen \& Ostriker 1992;
\markcite{sca93}Scaramella et al.\ 1993; \markcite{per95}Persi et
al. 1995). The resulting CMB fluctuations are expected to be of
order 1--10$\mu$K on 10 arcmin scales (see \S\ref{foregrounds_sz}).
A detailed analytical and numerical investigation of the SZ effect on
large scales will be presented in \markcite{ref98}Refregier et
al. (1998).

Because of the moderate density and temperature of the gas in
superclusters, the large-scale SZ effect is difficult to detect
directly. However, it could be detectable by cross-correlating the CMB
with optical galaxy catalogs which would act as tracers of the
density. Specifically, we can consider the galaxy-CMB correlation
function
\begin{equation}
\label{eq:wgc_def}
W_{gc}(\theta) \equiv \langle \frac{\Delta N_{g}}{\overline{N}_{g}} \Delta
T_{c} \rangle_{\theta},
\end{equation}
where $\frac{\Delta N_{g}}{\overline{N}_{g}}$ is the fluctation in the
number of galaxies and $\Delta T_{c}$ is the CMB temperature
fluctuation, in two pixels separated by an angle $\theta$.  A rough
estimate of the amplitude of $W_{gc}$ is given by
\begin{equation}
W_{gc}(\theta) \approx b_{\rm gas} \overline{\Delta T}_{\rm SZ}
W_{gg}(\theta),
\end{equation}
where $W_{gg}(\theta)$ is the galaxy-galaxy correlation function, and
$\overline{\Delta T}_{\rm SZ}$ is the SZ fluctuation amplitude for the
volume sampled by the galaxy catalog. The factor $b_{\rm gas}$ is the
bias between the gas pressure and the density traced by the galaxies.

To investigate the detectability of the correlation, let us consider
the measurement of the zero lag cross-correlation function $W_{gc}(0)$
by averaging over $N_{\rm pix}$ pixels in equation~(\ref{eq:wgc_def}).
In the null hypothesis corresponding to an absence of correlation
(i.e. $\overline{\Delta T}_{\rm SZ} = 0$), the fluctuation in
$W_{gc}(0)$ is $\sigma[W_{gc}(0)] = N_{\rm pix}^{-\frac{1}{2}}
\sigma[\frac{\Delta N_{g}}{\overline{N}_{g}}] \sigma[\Delta T_{c}]$,
where $\sigma[ Q ]$ indicate {\it rms} fluctuations in quantity Q.  The
galaxy count variance is the sum of a Poisson term and a clustering
term, i.e. $\sigma^{2}[\frac{\Delta N_{g}}{\overline{N}_{g}}] \equiv
\sigma_{g}^{2} = \overline{N}_{g}^{-1} + W_{gg}(0)$. The {\it rms} CMB
temperature results from the total ``noise'' produced by the CMB and
instrumental noise, i.e. $\sigma[\Delta T_{c}] = \Delta T_{\rm
noise}$, as in equation~(\ref{eq:deltaT_noise}). As a result, the
signal-to-noise ratio ${\rm SNR}_{gc} \equiv W_{gc}(0)/ \sigma[
W_{gc}(0) ]$ for measuring $W_{gc}(0)$ is given by
\begin{equation}
{\rm SNR}_{gc} \simeq b_{\rm gas} \overline{\Delta T}_{\rm SZ}
W_{gg}(0) N_{\rm pix}^{\frac{1}{2}} \sigma_{g}^{-1} \Delta T_{\rm
noise}^{-1}.
\end{equation}

For instance, the all-sky APM catalog (\markcite{mad90a}Maddox et
al. 1990a; \markcite{mad90b}Maddox, Efstathiou, \& Sutherland 1990b;
\markcite{irw92}Irwin \& McMahon 1992; \markcite{irw94}Irwin, Maddox,
\& McMahon\ 1994) contains $\sim 600$ galaxies per deg$^2$ with
E-magnitude less than about 19, thereby sampling redshifts $z\lesssim
0.2$. We consider $0\fdg3\times0\fdg3$ \hbox{MAP} pixels. For this
pixel size and magnitude cut, the zero-lag galaxy correlation function
for the APM catalog is $W_{gg}(0) \simeq 0.1$ (see measurement by
\markcite{mad90c}Maddox et al. 1990c). The resulting galaxy count {\it
rms} is then $\sigma_{g} \simeq 0.34$, and is dominated by the
clustering term.

The SZ amplitude $\overline{\Delta T}_{\rm SZ}$ is difficult to
estimate. The numerical simulations of \markcite{sca93} Scaramella et
al. (1993) yield a $y$-parameter integrated over all redshifts of $y(z
< \infty) \simeq 4 \times 10^{-7}$, for a standard CDM model with
$\sigma_{8} =0.67$ and after removing bright X-ray sources. This
corresponds to $\overline{\Delta T}_{\rm SZ}(z < \infty) \simeq
2.4\mu$K. About 25\% of $y$ is produced at $z \lesssim 0.2$ (see their
figure 9). This implies that the SZ amplitude for the APM survey is,
approximately, $\overline{\Delta T}_{\rm SZ}(z \lesssim 0.2) \simeq
0.6\mu$K. \markcite{per95}Persi et al.\ (1995) have studied a cluster
normalized $\Lambda$CDM model with $\Omega=0.45$, $h=0.65$ using
analytical techniques and numerical simulations. After removing bright
X-ray cores, they find {\it rms} temperature fluctuations equal to
$\overline{\Delta T}_{\rm SZ}(z < \infty) \simeq 6.0\mu$K, for the
above pixel size (see their figure 3). An integration of their figure
3a yields that about 46\% of these fluctuations are produced at $z
\lesssim 0.2$ . This implies that $\overline{\Delta T}_{\rm SZ}(z
\lesssim 0.2) \simeq 2.8\mu$K. We adopt $\overline{\Delta T}_{\rm
SZ}(z \lesssim 0.2) \approx 0.4 \times \overline{\Delta T}_{\rm SZ}(z
< \infty) \approx 2 \mu$K, as a compromise. Note that all the
temperatures in this paragraph are RJ temperatures in the $x
\rightarrow 0$ limit, and must be corrected by a factor of
$\frac{j(x)}{-2}$ to derive thermodynamic temperatures at higher
frequencies (see Eq.~[\ref{eq:deltat_sz}]).

With these numerical values, the signal-to-noise ratio for measuring
$W_{gc}(0)$ is given by
\begin{equation}
{\rm SNR}_{gc} \simeq 3.2 \left( \frac{b_{\rm gas}}{1} \right)
\left( \frac{\overline{\Delta T}_{\rm
SZ}}{2 \mu{\rm K}} \right) \left( \frac{W_{gg}(0)}{0.1} \right)
\left( \frac{75 \mu{\rm K}}{\Delta T_{\rm noise}} \right)
\left( \frac{0.34}{\sigma_{g}} \right)
\left( \frac{f_{\rm sky}}{0.7} \right)^{\frac{1}{2}}
\left( \frac{j(x)}{-1.56} \right) ,
\end{equation}
where $f_{\rm sky}$ is the fraction of the sky covered, and
$\overline{\Delta T}_{\rm SZ} \equiv \overline{\Delta T}_{\rm SZ}(z
\lesssim 0.2)$ in the RJ regime. The central values for the CMB noise
$\Delta T_{\rm noise}$ and the spectral function $j(x)$ were chosen to
be those relevant for the \hbox{MAP} 94 GHz channel after filtering
(Eqs.~[\ref{eq:deltaT_noise},\ref{eq:deltat_sz}]).  The correlation signal
is likely to be enhanced by a positive biasing of the gas pressure
($b_{\rm gas} > 1$), as is indicated from our preliminary study of the
clustering and shock heating of the hot IGM (\markcite{ref98}Refregier
et al.\ 1998). An APM-\hbox{MAP} cross-correlation should thus yield a
marginal detection, or at least an interesting upper-limit of the SZ
effect by large scale structure. Note that the use of the Sloan
Digital Sky Survey (\markcite{gun93}Gunn \& Knapp 1993) with
photometric redshifts could yield increased significance and,
possibly, redshift information.

It is instructive to express these results in terms of constraints on
the Compton $y$-parameter. In the absence of a detectable
correlation, the above expression yields a $3\sigma$ upper
limit on the {\it rms} Compton parameter of
\begin{equation}
\delta y(z\lesssim 0.2) \lesssim 3.4 \times 10^{-7}.
\end{equation}
The $3\sigma$ upper-limit on the total integrated y-parameter (see
Eq.~[\ref{eq:def_y}]) set by the COBE-FIRAS measurement
(\markcite{fix96}Fixsen et al.\ 1996) of the spectral distortion of
the CMB is $y(z<\infty) < 2.2 \times 10^{-5}$. The $3\sigma$ upper
limits derived from a cross-correlation of DMR and FIRAS on board COBE
is $\delta y(z<\infty) < 4.5 \times 10^{-6}$ (\markcite{fix97}Fixsen
et al.\ 1997). As discussed above, one expects $y(z \lesssim .2)
\approx 0.4 \times y(z < \infty)$.  The APM-\hbox{MAP} correlation
limits thus provide an improvement by a factor of about 26 and 5 over
these previous analyses, respectively.

Our expected upper limit can also be compared with previous attempts
to detect a cross-correlation between the CMB and tracers of the
large-scale structure. \markcite{ban96}Banday et al.\ (1996; see also
\markcite{ben93}Bennett et al.\ 1993) have set a $3\sigma$ upper limit
of $\delta y(z \lesssim 0.2) < 1.5 \times 10^{-6}$ by
cross-correlating COBE-DMR 4-year maps with the ACO cluster
catalog. This is a factor of about 4 larger than our expected
$3\sigma$ upper limit. Other limits have been derived from the
cross-correlation of the CMB with the X-ray Background (XRB), a
fraction of which could be produced by bremsstrahlung emission from
the hot gas also responsible for the SZ effect. From a
cross-correlation of COBE with the HEAO-1 X-ray map, Banday et
al. find $\delta y < 1.3 \times 10^{-6}$ at the $3\sigma$ level (see
also \markcite{bou93}Boughn \& Jahoda 1993). A similar analysis
between COBE and the ROSAT All Sky Survey yielded a $3\sigma$ limit of
$\delta y < 1.2 \times 10^{-6}$ (\markcite{knei97}Kneissl et al.\
1997). The redshift range for the CMB-XRB correlations is however hard
to estimate, making these limits difficult to compare with our
expected limits.

Constraints on the hot gas on supercluster scales would not only
benefit primordial CMB measurements, but would also be of
astrophysical importance. It is indeed well known that stars and
galaxies contain at most 20\% of the baryonic mass predicted by the
theory of Big Bang Nucleosynthesis. The remaining $\sim80$\%, the
``missing baryons'', are most likely to be in the form of the hot gas
in groups, clusters and superclusters (\markcite{Fuk97}Fukugita,
Hogan, \& Peebles 1997).  While the X-ray emission of the hot IGM in
clusters is now well established, the hot gas in structures with
moderate density constrasts such as groups and supercluster filaments
is difficult to detect. It is indeed too cold to be detectable in the
X-ray band, and too hot to be observable as quasar absorption lines. A
cross-correlation of the CMB with galaxy catalogs might thus be the
only sensitive probe of the missing baryons.

\section{Conclusions}
\label{conclusion}
While the primary anisotropies produced at the surface of last
scattering dominate CMB fluctuations on large angular scales, various
extragalactic foregrounds can be important on 10 arcminute
scales. Upon surveying the major extragalactic foregrounds, we showed
that discrete sources, the SZ effect, and gravitational lensing are
the dominant extragalactic foregrounds for the \hbox{MAP}
experiment. We estimated that \hbox{MAP} will detect about 46 discrete
sources directly at the $>5\sigma$ confidence level. The residual
undetected sources will produce a detectable ($\sim 5 \sigma$)
skewness which should not be mistaken for that of intrinsic CMB
fluctuations. \hbox{MAP} will also detect directly about 10 SZ
clusters at the above confidence level. In particular, \hbox{MAP} will
clearly detect and somewhat resolve the SZ decrement from the Coma
cluster.  We also estimated that about 80\% of the brightest SZ
clusters are contained in the XBACs cluster catalog, and thus already
have known positions, redshifts and X-ray properties.

Fainter extragalactic fluctuations can be probed statistically by
cross-correlating CMB maps with external catalogs. In particular, a
cross-correlation of \hbox{MAP} with cluster positions from the XBACs
catalog will probe SZ clusters with $S_{94} \gtrsim 200$ mJy, an order
of magnitude fainter than the $5 \sigma$ threshold for direct
detection by \hbox{MAP}. This will provide a test of the virialization
state of clusters and a measurement of their gas fractions. A
cross-correlation of \hbox{MAP} with existing galaxy catalogs will
permit to probe the hot gas on supercluster scales. In particular, we
estimated that a cross-correlation of \hbox{MAP} with the APM galaxy
survey will yield a marginal detection, or at least a tight upper
limit on SZ fluctuations which are correlated with large-scale
structure. Assuming that the hot gas follows galaxy counts, this upper
limit would be a four-fold improvement on the COBE upper limits on the
{\it rms} Compton $y$-parameter. This would provide constrains on the
``missing baryons'', a large fraction of which is likely to be in the
form of hot gas in groups and supercluster filaments and is difficult
to detect by other means.

\acknowledgments We thank Norman Jarosik and Lyman Page for their help
with \hbox{MAP} specifications. We are also grateful to Neta Bahcall,
Michael Strauss, Chuck Bennett and Siang Peng Oh for useful
discussions, and to Hans B\"{o}hringer for providing us with the X-ray
parameters for Virgo. We also thank the editor, Ned Wright, and an
anonymous referee for a careful review of the manuscript, and Eric
Gawiser and Yasushi Suto for useful comments.  This work by supported
by the MAP MIDEX program and the NASA ATP grant NAG5-7154.

{}

\begin{deluxetable}{rrrrrr}
\tablecaption{Summary of Extragalactic Foregrounds for $\nu=94$
GHz and $\ell=450$.
\label{tab:foregrounds}}
\tablewidth{0pt}
\tablehead{
\colhead{Source} &
\colhead{$\Delta T_{\ell}$ ($\mu$K)\tablenotemark{a}} &
\colhead{$\ell_{\rm eq}$\tablenotemark{b}} &
\colhead{Thermal\tablenotemark{c}} & \colhead{Note} &
\colhead{Ref.\tablenotemark{d}}
}
\startdata
CMB\tablenotemark{e} & 50 & & yes & & 1\\
Discrete\tablenotemark{f} & 5   & 1800      & no & $S<1.0$  Jy  & 2 \\
                          & 2   & 3100      & no & $S<0.1$ Jy     & 2 \\
SZ\tablenotemark{g} & 10   & 1900      & no & C     &3 \\
                    & 7    & 2300      & no & NC    &3 \\ 
OV\tablenotemark{h} & 2   & 2900      & yes & $z_{r}=50$    & 4 \\
                    & 1   & 3100      & yes & $z_{r}=10$    & 4 \\
ISW & 1     & 5000      & yes & $\Omega h=0.25$     & 5 \\
    & 0.9   & 5800      & yes & $\Omega h=0.50$      & 5 \\
Lensing & 5 & 2400 & yes & & 6\\
\enddata
\normalsize
\tablenotetext{a}{$\Delta T_{\ell} \equiv [\ell(2\ell +1)C_{\ell}/4\pi]^{1/2}$
centered at $\ell=450$ and $\nu=94$ GHz.}
\tablenotetext{b}{Value of $\ell$ for which $\Delta T_{\ell}=\Delta
T_{\ell,{\rm CMB}}$}
\tablenotetext{c}{Thermal (yes) or nonthermal (no) spectral dependence}
\tablenotetext{d}{1: Seljak \& Zaldarriaga 1996; 2: Toffolatti
et al.\ 1998; 3: Persi et al.\ 1995; 4: Hu \& White 1996;
5: Seljak 1996a; 6: Zaldarriaga \& Seljak 1998}
\tablenotetext{e}{Primordial CMB fluctuations for a CDM model
with $\Omega_{m}=1$, $\Omega_b=0.05$, and $h=0.5$}
\tablenotetext{f}{Discrete sources with 94 GHz removal threshold of
0.1, 1 Jy, respectively}
\tablenotetext{g}{SZ effect with (C) and without (NC) cluster cores
respectively.}
\tablenotetext{h}{OV effect with two different reionization redshifts
$z_{r}$}
\end{deluxetable}

\begin{deluxetable}{rlrrrrr}
\tablecaption{Brightest SZ clusters in the XBACS catalog
\label{tab:clusters}}
\tablewidth{0pt}
\tablehead{
\colhead{Rank} & \colhead{Name} &
\colhead{$z$\tablenotemark{a}} &
\colhead{$kT_{e}$ (keV)\tablenotemark{b}} &
\colhead{$S_{94}$ (Jy)\tablenotemark{c}} &
\colhead{SNR\tablenotemark{d}} & \colhead{Ref.\tablenotemark{g}} }
\startdata
  1 & A1656 (Coma)      & 0.023  &  8.1  & 10.77 & 26.4\tablenotemark{f} & 1 \\
  2 & A3526 (Centaurus) & 0.011  &  3.9  &  7.80 & 19.1 & 1 \\
  3 & A 426 (Perseus)   & 0.018  &  5.5  &  6.46 & 15.8 & 1 \\
  4 & A1060  (Hydra)    & 0.012  &  3.3  &  3.85 &  9.4 & 1 \\
  5 & A3627  & 0.016  &  4.0\tablenotemark{e} &  3.74 &  9.2 & 2 \\
  6 & A3571  & 0.039  &  7.6  &  3.43 &  8.4 & 1 \\
  7 & A2319  & 0.056  &  9.9  &  3.36 &  8.2 & 1 \\
  8 & A 754  & 0.053  &  9.1  &  3.07 &  7.5 & 1 \\
  9 & A2199  & 0.030  &  4.7  &  1.70 &  4.2 & 2 \\
 10 & A2256  & 0.058  &  7.5  &  1.59 &  3.9 & 1 \\
 11 & A1314  & 0.034  &  5.0  &  1.57 &  3.9 & 1 \\
 12 & A1367  & 0.021  &  3.5  &  1.54 &  3.8 & 1 \\
\enddata
\normalsize
\tablenotetext{a}{Cluster redshift}
\tablenotetext{b}{Electron temperature determined from X-ray
observations}
\tablenotetext{c}{SZ flux at 94 GHz estimated assuming virialization
and h=0.5}
\tablenotetext{d}{Signal-to-noise ratio for the detection of an
unresolved cluster with the \hbox{MAP} 94 GHz channel}
\tablenotetext{e}{X-ray temperature estimated from the 
luminosity-temperature relation}
\tablenotetext{f}{SNR overestimated since Coma will be resolved}
\tablenotetext{g}{Reference for $z$ and $kT_{e}$:
1: David et al.\ 1993; 2: Ebeling et al.\ 1996.}
\end{deluxetable}

\end{document}